\documentclass[12pt]{article}
\usepackage{latexsym, epsfig, graphics}
\newcommand{\be}{\begin{equation}}
\newcommand{\ee}{\end{equation}}
\newcommand{\bq}{\begin{eqnarray}}
\newcommand{\eq}{\end{eqnarray}}
\newcommand{\intsigt}{\int_{0}^{p^{+}} d\sigma \int_{\tau_{i}}^{\tau_{f}}
d\tau}
\begin{document}
\begin{titlepage}
\today          \hfill 
\begin{center}
\hfill    LBNL-56805 \\
         \hfill    NSF-KITP-05-02 \\

\vskip .5in

{\large \bf Further Results About Field Theory On The World Sheet
And String Formation}
\footnote{This work was supported in part
 by the Director, Office of Science,
 Office of High Energy and Nuclear Physics, 
 of the U.S. Department of Energy under Contract 
DE-AC03-76SF00098, and in part by the National Science Foundation Grant
No.PHY99-07949}
\vskip .50in


\vskip .5in
Korkut Bardakci

{\em Department of Physics\\
University of California at Berkeley\\
   and\\
 Theoretical Physics Group\\
    Lawrence Berkeley National Laboratory\\
      University of California\\
    Berkeley, California 94720}
\end{center}

\vskip .5in

\begin{abstract}
The present article is the continuation of the earlier work, which
used the world sheet representation  and the mean field
approximation to sum planar graphs in massless $\phi^{3}$
field theory. We improve on
the previous work in two respects: A prefactor in the world sheet
 propagator that had been neglected is now taken into account.
In addition, we introduce a non-zero bare mass for the field $\phi$.
Working with a theory with cutoff and using the mean field
approximation, we find that, depending on the range of values of the 
mass and coupling constant, the model has two phases: A string forming
 phase and a perturbative field theory phase. We also find the generation
of a new degree of freedom, which was not in the model originally. This
new degree of freedom can be thought of as the string slope, which is
now promoted into a fluctuating dynamical variable. Finally, we show
that the introduction of the bare mass makes it possible to
renormalize the model. 

\end{abstract}
\end{titlepage}

\newpage
\renewcommand{\thepage}{\arabic{page}}
\setcounter{page}{1}
\noindent{\bf 1. Introduction}

\vskip 9pt

The present article is the continuation of a series of articles [1,2,3,4,5]
pursuing a program of summing planar graphs in field theory. Because
of its simplicity, the theory most intensively investigated so far is the
$\phi^{3}$ theory, although progress has been made in extending the
program to more physical theories [6,7,8]. The basic idea, due
 to 't Hooft [9],
is to represent planar Feynman graphs on the world sheet, using a mixture
of position and momentum space light cone variables as coordinates. This
representation, which was originally non-local, was later reformulated
as a local field theory on the world sheet by introducing additional
non-dynamical fields [1]. The advantage of this
 reformulation is that it provides a useful setup for studying the sum
of planar Feynman graphs, since one can then appeal to various approximation
schemes familiar from field theory to investigate problems of interest.

The problem we are going to investigate in this article is string
formation in the $\phi^{3}$ field theory and the approximation scheme
we are going to use is the mean field method. Although the mean field
method has its limitations, we hope that at least the predictions
it makes about basic dynamical questions such as string formation are
qualitatively correct. In any
case, it is a simple method to use and we have nothing better available
at the present.

The present paper can be viewed as a follow up to reference [5].
As such, it has a good deal of overlap with it,
as well as with some of the earlier work on the same subject. This is
because in organizing this paper, the goal was to present a self
contained treatment, which should be intelligible even to a reader
unfamiliar with the previous work on the subject. When we preview
the rest of the paper below, we will try to make clear what is new
and what is a review of the earlier work  cited at the
beginning of this section.

Before getting started, it may be helpful to summarize the advances
made  in this paper in comparison to [5]. Apart from some
simplification and streamlining of the treatment, there are two new
features of interest. The first one is somewhat technical: The
prefactor $1/(2 p^{+})$ in eq.(2) for the propagator
 was neglected in the previous applications of the mean field method.
In the present work, this factor is taken into account. This does
effect the details of calculations, but it does not qualitatively change
the final results, concerning, for example, string formation.

The second new feature is more significant. In reference [5], as well as
in the work preceding it, the field theoretic bare mass was taken 
to be zero. Of course, in any case, a non-zero mass is generated in
higher orders of perturbation theory, which is cutoff dependent
and needs renormalization. This cutoff dependence shows up in the
expression for the ground state energy (eq.(56)), which is proportional
to the cutoff parameter $1/a^{2}$.
In the present work, we show how to introduce a non-zero
bare mass, which can then be used as a counter term to cancel the cutoff
dependent part of the ground state energy. 

We would  like to draw the attention of the reader to the main results
obtained in this article. We show that, in the cutoff theory
prior to renormalization, there is string formation for large enough
values of the coupling constant, whereas for smaller values of the
coupling constant, the model is in the perturbative field theory
phase. We  also show that physical quantities can be
made cutoff independent by introducing a suitable bare mass. From the
string perspective, the physical quantities we are referring to are
the string slope and the intercept. With the introduction of mass,
there is again string formation for a range of values of mass and
coupling constant, and field theory phase takes over when the parameters
are outside this range.
An important problem not addressed in this paper is whether string
formation observed in the cutoff theory persists after renormalization.

In section 2, we briefly review both the rules for Feynman graphs
in the light cone variables [9] and the local field theory on the
world sheet which generates these graphs [1]. We also discuss the 
transformation properties of the fields under a special Lorentz boost,
which manifests itself as a scale transformation on the world sheet. Since
lack of invariance under this scaling would imply violation of
Lorentz invariance, as we go along, we make sure that no such violation
occurs. This is an abridged version of a more complete discussion given
in [4].

In the local field theory discussed in section 2, the boundary
conditions on the fields were imposed by hand. In section 3, a more
general field theory is constructed, where the boundary conditions
are enforced by means of Lagrange multipliers. In addition, auxilliary
fermionic fields that distinguish between the boundaries and the bulk
are needed. To have a well defined theory, we have to introduce two
distinct cutoffs; one associated with the coordinate $\tau$ and the
other with $\sigma$. Until section 8, we will be working exclusively
with the cutoff theory, and all the results obtained will refer to 
this theory. The motivation for studying the cutoff theory is that
it is an interesting model in its own right and also it is an
indispensible preliminary to renormalization.
 This section follows reference [5] closely,
with the exception that, unlike in [5], we do not impose supersymmetry
on the model.
Although the use supersymmetry is an elegant way of taking care of
the matter-ghost cancellations, in retrospect, ghosts do not contribute
in any significant way to the meanfield calculations.
In the interests brevity and simplicity, we have therefore
 decided to drop the ghost sector and write down a non-supersymmetric model.

The mean field method which is at the basis of the present work
is discussed in section 4 from the point of view of the large $D$ 
limit, where $D$ is the dimension of the transverse space. This section
is largely a review of the material developed in the earlier work.
The only thing new is a brief comparison of the determinant resulting
from the integration over matter fields with the corresponding result
well known from string theory.

In section 5, the fermionic part of the action is diagonalized, and
the two fermionic energy levels are calculated as a function of a parameter
 $x$, which serves as an order parameter.
 It is argued that, independent of any approximation
scheme, a non-zero order parameter signals both string formation and
condensation of Feynman graphs on the world sheet, thereby linking
these two phenomena. On the other hand, $x=0$ corresponds to the
perturbative field theory phase of the model.
 This section is a mostly a review of the material
 covered in the previous work.

In section 6, the prefactor $1/(2 p^{+})$ in the propagator (eq.(2)),
neglected in the previous work, is taken into account in the leading
order of the mean field approximation. As a result, the interaction
vertex becomes a function of the order parameter $x$ (eq.(52)). The
material covered in this section is completely new.

In section 7, combining the results of the previous sections,
the ground state energy is expressed in terms of $x$. Minimizing
this energy, we find that there is a critical value of the coupling
constant, $g=g_{c}$: For $g<g_{c}$, the minimum is at $x=0$, whereas
for $g>g_{c}$, the minimum is at some $x\neq 0$. It then follows that,
at $g=g_{c}$, there is a transition from the perturbative field theory
phase to the string phase. A considerable portion of this section 
is new.

So far, we have been studying a model where the field theory bare
mass was set equal to zero. In section 8, we show how to introduce
a non-zero bare mass term in the leading order of the mean field
approximation, and we show that, if the mass is not too large, 
 the same picture as in the massless case emerges: There is
a $g_{c}$, dependent on the mass, that seperates the string forming
and perturbative phases of the model. It is also possible to use
the bare mass as a counter term to eliminate the cutoff dependence
of the ground state energy. From the string perspective, this
corresponds to the renormalization of the intercept. Since, in section
5, we have already shown that the string slope is already finite, at least
for a free string, this is all the renormalization that is needed. The
investigation of the properties of the renormalized theory, which could
be quite different from those of the cutoff theory studied so far, is
left for future research. This section is completely new.

All the results obtained so far were in the leading order of the mean field
approximation, or what is the same thing, in the leading order of large
$D$ limit. In section 9, we compute a particular contribution coming
from the next order of the large $D$ limit. This contribution
is important in that it provides
the kinetic energy term for a new propagating degree of freedom which
was not present in the original action. As a result, the string slope,
which was a constant in the leading order, becomes a dynamical field
which can fluctuate around its mean value. We end the section with some
speculative remarks about the possible connection between this new 
dynamical field and the extra dimension in AdS/CFT correspondence [10,11],
and how it could possibly interpolate between ``hard'' and ``soft''
high energy processes.
This section largely overlaps with the corresponding material in [5];
it is included here for the sake of completeness.

Finally, Appendices A and B contain some of the additional details of
 the mean field calculations developed in sections 6 and 8.

\vskip 9pt
\noindent{\bf 2. A Brief Review}
\vskip 9pt
The Feynman graphs of a massive $\phi^{3}$ theory have a particularly
simple form in the mixed lightcone representation of 't Hooft [9]. In
this representation, the evolution parameter, $\tau$, is $x^{+}$, and the
conjugate Hamiltonian is $p^{-}$. The Minkowski evolution operator is 
given by
\be
T=\exp(-i x^{+} p^{-}).
\ee
The notation is such that a Minkowski four vector $v^{\mu}$ has the light
cone components $(v^{+},v^{-},{\bf v})$, where $v^{\pm}=(v^{0}\pm v^{1})/
\sqrt{2}$, and boldface letters label the transverse directions. A propagator
that carries momentum $p$ is pictured as a horizontal strip of width $p^{+}$
and length $\tau=x^{+}$, bounded by two solid lines on the world sheet
(Fig.1).
\begin{figure}[t]
\centerline{\epsfig{file=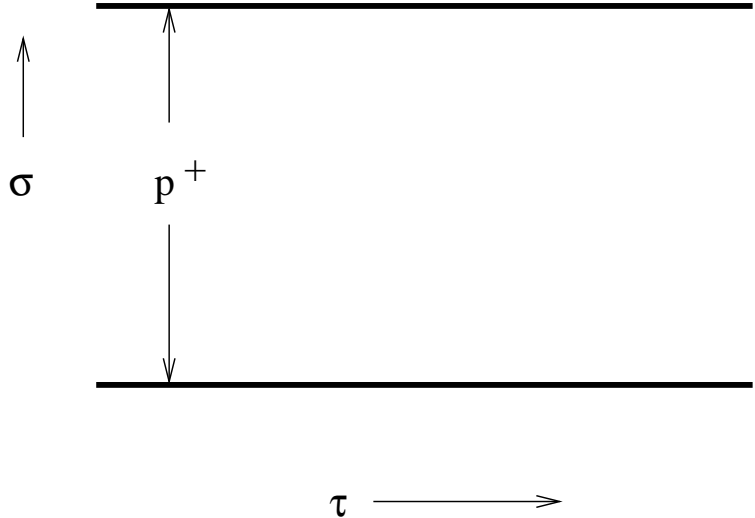,width=10cm}}
\caption{The Propagator}
\end{figure}
 The solid lines forming the boundary carry transverse momenta
${\bf q}_{1}$ and ${\bf q}_{2}$, with
$$
{\bf p}={\bf q}_{1}-{\bf q}_{2},
$$
and the corresponding propagator is given by
\be
\Delta(p)=\frac{\theta(\tau)}{2 p^{+}}\,\exp\left(-i\,\frac{\tau}{2 p^{+}}
({\bf p}^{2}+m^{2})\right).
\ee

More complicated graphs consist of several horizontal solid line segments
(Fig.2).
\begin{figure}[t]
\centerline{\epsfig{file=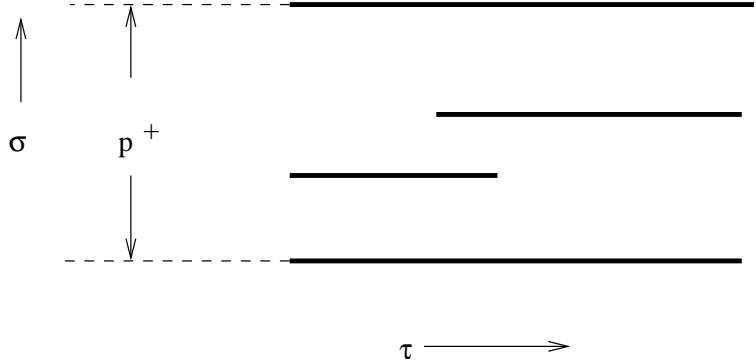,width=10cm}}
\caption{A Typical Graph}
\end{figure}
 A factor of $g$, the coupling constant, is associated with the
beginning and the end of each line segment, where the interaction takes
place. Finally, one has to integrate over the positions of the interaction
vertices, as well as the momenta carried by the solid lines. A typical light
cone graph is pictured in Fig.2.

It was shown in [1] that the light cone Feynman rules described above can be
reproduced by a local world sheet field theory. The world sheet is
 parametrized
by the coordinates $\sigma$ along the $p^{+}$ direction and $\tau$ along the
$x^{+}$ direction, and the transverse momentum ${\bf q}$ is promoted to a
 bosonic
field ${\bf q}(\sigma,\tau)$ on the world sheet. In addition, two fermionic
fields (ghosts) $b(\sigma,\tau)$ and $c(\sigma,\tau)$ are needed. In contrast
 to
${\bf q}$, which has D components, $b$ and $c$ each have $D/2$ components.
 Here,
D is the dimension of the transverse space, assumed to be even. Here and in
 the next section, we will consider first consider the zero mass case for the
sake of simplicity, and postpone the discussion of the the massive theory to
section (8). The action on
 the world sheet for the massless theory($m=0$), is given by
\be
S_{0}=\intsigt\left(b'\cdot c'-\frac{1}{2}\,{\bf q}'^{2}\right),
\ee
where the prime denotes derivative with respect to $\sigma$. This action is 
supplemented by Dirichlet boundary conditions
\be
\dot{{\bf q}}=0,\; b=c=0,
\ee
on the solid lines, where the dot denotes derivative with respect to $\tau$. It
was shown in [1] that if the equations of motion for ${\bf q}$ are solved
subject to the above boundary conditions, the resulting classical action
reproduces the exponential factor in eq.(2). The unwanted quantum contribution
$$
- \frac{D}{2} \det(\partial_{\sigma}^{2})
$$
is cancelled by the corresponding determinant resulting from integrating
over ghosts.

The action formulation described above was extensively used in the previous
work [2,3,4,5]. It has, however, two defects: The prefactor $1/(2 p^{+})$
in eq.(2) is missing and a non-zero bare mass cannot be accomodated. Since
there is no symmetry forcing the renormalized mass to be zero, this
means that the mass counter term needed for renormalization cannot be
introduced. In sections (6) and (8), we will show how to overcome both of
these problems in the context of the mean field approximation.

Finally, we would like to discuss briefly the question of Lorentz
invariance. This is a non-trivial problem, since the use of the light cone
variables obscures the Lorentz transformation properties of the fields.
There is, however, a special subgroup of the Lorentz group, under which
the light cone coordinates have simple linear transformation properties.
If $L_{i,j}$ are the angular momenta and $K_{i}$ are the boosts, the
generators of this subgroup are
\be
L_{i,j},\;\; M_{+,-}=K_{1},\;\; M_{+,i}=K_{i}+L_{1,i},
\ee
where the indices $i$ and $j$ run from 2 to $D+2$. It turns out that invariance
under all the generators, with the sole exception of $K_{1}$, is rather trivial
[4]. The non-trivial transformation generated by $K_{1}$ corresponds to
scaling of $x^{+}$ and $p^{+}$ by a constant u:
\be
x^{+}\rightarrow x^{+}/u,\;\; p^{+}\rightarrow p^{+}/u,
\ee
leaving the transverse momenta ${\bf q}$ unchanged. As we go along, we will
check the invariance the invariance of our equations and our results
under this scale transformation. However, in this paper we will not
investigate the problem of invariance under the full Lorentz group \footnote{
The problems of Lorentz invariance and renormalization are closely related.
See [12] for an investigation of both problems in the light cone formalism.}

Let us check the scale invariance of (2) and (3). In eq. (2), the
 exponential is clearly invariant, and the prefactor $1/(2 p^{+})$ is
the integration measure that makes integration over $p^{+}$ invariant.
The action (3) and the boundary conditions (4) are also invariant
 if the fields transform according to
\be
{\bf q}(\sigma,\tau)\rightarrow {\bf q}(u\sigma,u\tau),\;\;
b,c(\sigma,\tau)\rightarrow b,c(u\sigma,u\tau).
\ee
\vskip 9pt

\noindent{\bf 3. The Complete World Sheet Action For m=0}
\vskip 9pt

It is possible to include the boundary conditions of
eq.(4) in the action itself,
rather then imposing them by hand [2,3,4]. What follows is a condensed
version of the treatment given in [3,4]. The complete world sheet action in the
 zero mass case can be written as a sum of four terms:
\be
S= S_{m}+ S_{g}+ S_{g.f}+ S_{f}.
\ee
$S_{m}$, the matter action, and $S_{g}$, the ghost action, are given by 
\bq
S_{m}&=&\intsigt \left(-\frac{1}{2} {\bf q}'^{2}+ \rho\, {\bf y}\cdot
\dot{{\bf q}}\right),\nonumber\\
S_{g}&=&\intsigt \left(b'\cdot c'+ \rho\, \bar{b}\cdot b+ \rho\, \bar{c}
\cdot c\right).
\eq
Here the boundary conditions on the matter and ghost fields are implemented
by means of the Lagrange multiplier fields ${\bf y}$, $\bar{b}$ and $\bar{c}$.
The field $\rho$ is a delta function on the boundaries(solid lines) and zero
in the bulk, in order to ensure that the boundary conditions are imposed
on the boundaries and not in the bulk.
 We will shortly give an explicit construction for $\rho$
in terms of fermionic fields. $S_{g.f}$ is the gauge fixing term given by
\be
S_{g.f}=\intsigt\left(-\frac{1}{2} \bar{\rho}\, \alpha^{2}
 {\bf y}^{2}\right),
\ee
where $\alpha$ is a constant and
 $\bar{\rho}$ is complementary to $\rho$; it vanishes on the boundaries
and is equal to one everywhere else. In the absence of this term,
the integration over over ${\bf y}$ the action
is invariant under a gauge transformation of the form
$$
{\bf y}\rightarrow {\bf y}+ {\bf z} \bar{\rho},
$$
where ${\bf z}$ is an arbirary function of the coordinates. This gauge 
invariance causes the functional integral over ${\bf y}$
 to diverge in the bulk where $\bar{\rho}$ is zero;
 fixing the gauge eliminates this divergence. There is also
a Faddeev-Popov type measure factor associated with gauge fixing. One can see
the need for it as follows: The
integral over ${\bf y}$ is Gaussian away from the boundaries  and it can
explicitly be evaluated, resulting in a singular contribution that depends
on the gauge fixing parameter $\alpha$. If we regulate this singular
 expression by discretizing the $\tau$ coordinate into segments of length
$a'$, we have a product of the form
$$
\prod (a' \alpha^{2})^{- D/2}.
$$
This is an unwanted factor resulting from gauge fixing. It can be cancelled
by introducing a compensating measure factor in the ${\bf y}$, or by choosing
\be
a' \alpha^{2}=1.
\ee
From now on, we will assume that $\alpha$ is fixed by this relation.
We would like to point out that
 in the earlier work, for example reference [5], $\alpha$
was left arbitrary. It turns out that as a consequence of eq.(28), the
string slope parameter, eq.(42) comes out finite without any need of
renormalization.

It remains to specify the fermionic action $S_{f}$. Introducing a two
component fermion field $\psi_{i}(\sigma,\tau)$, $i=1,2$, and its adjoint
$\bar{\psi}_{i}(\sigma,\tau)$, $\rho$ and $\bar{\rho}$ of eqs.(9,10) can be
expressed as 
\be
\rho=\frac{1}{2}\bar{\psi}(1-\sigma_{3})\psi,\;\;
\bar{\rho}=\frac{1}{2}\bar{\psi}(1+\sigma_{3})\psi.
\ee
To see how this works, it is best to discretize the $\sigma$ coordinate
into segments of length $a$, which also helps regulate the fermionic
sector. This discretization is pictorially represented in Fig.3 by
horizontal lines at constant $\sigma$, spaced distance $a$ apart.
\begin{figure}[t]
\centerline{\epsfig{file=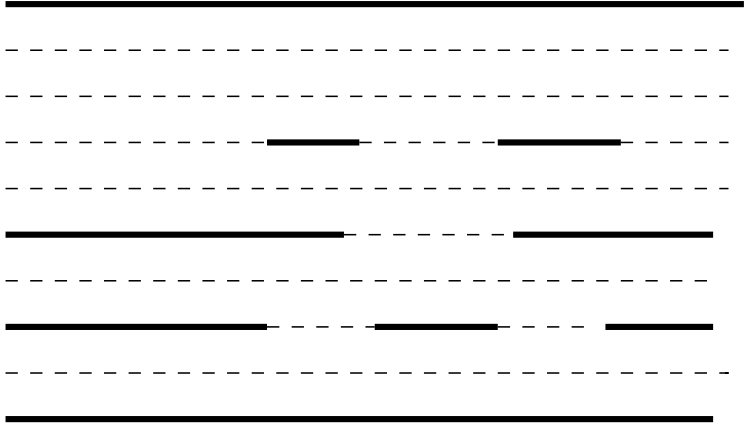,width=10cm}}
\caption{Solid and Dotted Lines}
\end{figure}
 The boundaries are marked by solid lines as before, and the rest of the
space (the bulk) is filled with dotted lines. If we identify the spin up 
component of the fermionic field ($i=1$) with the dotted lines and the
spin down component ($i=2$) with the solid lines, $\rho$ and $\bar{\rho}$
are then the spin down and spin up projection operators. Therefore,
$\rho=1$ on solid lines and it is zero on dotted lines, as stated before, and
the opposite holds for $\bar{\rho}$.

We have so far introduced  two cutoff parameters, $a$ and $a'$. All the
 results obtained until section 8 refer to the cutoff model regularized
by these two parameters. In section 8, we will show how the cutoff
dependence can be eliminated by means of mass renormalization.
Unfortunately, in this paper, we will not have much to say about the
renormalized model.

 With these preliminaries out of the way,
the fermionic action is given by
\bq
S_{f}&=&\sum_{n}\int d\tau\left(i \bar{\psi}_{n} \dot{\psi}_{n}
-\tilde{g} D\, \bar{\psi}_{n} \sigma_{1}\psi_{n}\right)\nonumber\\
&\rightarrow& \intsigt \left(i \bar{\psi} \dot{\psi} - \tilde{g} D\,
\bar{\psi}\sigma_{1}\psi\right).
\eq
The first line is the action in terms of discretized fermions
$$
\psi_{n}(\tau)=\psi(n a,\tau),
$$
and the second line is in terms of continuum fermions. Since the canonical
anticommutation relations  are in terms of a Kroenecker delta in the first
case and a Dirac delta in the second case, these two
fermions differ in normalization by a factor of $\sqrt{a}$. The first term
in the action represents the time propagation of the free fermion
without spin flip, the solid and dotted lines propagating unchanged. The second
term, in flipping the spin, converts a solid line into a dotted line and
vice versa. This spin flip represents the $\phi^{3}$ interaction and it is
 accompanied by the coupling constant $\tilde{g}$, which we have
scaled by a factor of D for later convenience. As it will become clear later,
this is necessary to have a non-trivial large $D$ limit; otherwise, the theory
would be non-interacting in this limit. Note that, with this scaling, large
$D$ is the same as strong coupling. 
Now if $\tilde{g}$ is taken to 
be a constant, as was done in the earlier work [3,4,5], then the world sheet
field theory would reproduce only the exponential in eq.(2), and the factor
$1/(2 p^{+})$ would be missing. Later, we will show how to take care of
this factor in the mean field approximation by allowing it to become a
function of the dynamical variables.

We will now rewrite the full action, collecting the terms for $S_{m}$,
$S_{g.f}$ and $S_{f}$, but excluding $S_{g}$. We will argue later that
the inclusion of $S_{g}$ does not affect the dynamics of the problem.
 With this omission, $S$ is given
by
\bq
S&=&\intsigt\Big(-\frac{1}{2} {\bf q}'^{2}+ i\bar{\psi}\dot{\psi}
-\tilde{g} D\, \bar{\psi}\sigma_{1}\psi+\frac{1}{2}\, {\bf y}\cdot
 \dot{{\bf q}}
\,\bar{\psi}(1-\sigma_{3})\psi\nonumber\\
&-&\frac{1}{4}\alpha^{2} {\bf y}^{2} \bar{\psi}(1+\sigma_{3})\psi\Big).
\eq
Let us check the scale invariance of this action. If under scaling,
 the fields ${\bf q}$, ${\bf y}$  and the fermions  transform as
\bq
{\bf q}(\sigma,\tau)&\rightarrow&{\bf q}(u\sigma,u\tau),\;\;
{\bf y}(\sigma,\tau)\rightarrow {\bf y}(u\sigma,u\tau),\nonumber\\
\psi(\sigma,\tau)&\rightarrow& \sqrt{u}\psi(u\sigma,u\tau),\;\;
\bar{\psi}(\sigma,\tau)\rightarrow\sqrt{u}\bar{\psi}(u\sigma,u\tau),
\eq
 then all the terms are invariant
 except the interaction term proportional to
$\tilde{g}$ and the gauge fixing term proportional to $\alpha^{2}$.
 These two terms become invariant only by demanding that $\tilde{g}$
and $\alpha^{2}$ transform by
\be
\tilde{g}\rightarrow u \tilde{g},\;\;
\alpha^{2}\rightarrow u \alpha^{2}.
\ee

We will eventually fix $\alpha^{2}$ by setting $a' \alpha^{2}=1$, as
in eq.(11). This brings up the question whether the lattice spacings $a$
and $a'$ in the $\sigma$ and  $\tau$ directions transform under scaling. 
It is unusual to assign transformation properties to a cutoff; however,
we will argue that in this case it is quite natural. For example, if we
split the interval from $\sigma=0$ to $\sigma=p^{+}$ in N segments of length
$a$, it is clear that under scaling, N, being an integer, does not
change, and therefore, $a$ must transform like $p^{+}$. A similar
argument applies to $a'$, so we must have
\be
a\rightarrow a/u,\;\;a'\rightarrow a'/u
\ee
under scaling. This shows that $\alpha^{2}$, fixed by eq.(11), scales
correctly. We will see that the same is true for $\tilde{g}$ in section (6).

\vskip 9pt

\noindent{\bf 4. The Meanfield Approximation}

\vskip 9pt

The meanfield method as applied to this problem was developed in [3] and [5].
We will mainly follow the treatment given in [5], identifying the mean
field method with the large $D$ limit. Unlike in [5] however, there will be
no supersymmetry on the world sheet. We notice that the action (14)
represents a vector model, which can be solved in the large $D$ limit [13].
The standard approach is to replace the scalar products of  the vector
fields ${\bf y}$ and ${\bf q}$, namely
 ${\bf y}\cdot \dot{{\bf q}}$ and ${\bf y}^{2}$, by their vacuum
expectation values. The functional integral over the remaining fields is
carried out exactly, and the resulting effective action is minimized
with respect to the vacuum expectation values. An efficient way of carrying
out this program is to introduce two composite fields $\lambda_{1}$ and
${\lambda_{2}}$ by adding a term $\Delta S$ to the action:
\bq
S &\rightarrow& S+ \Delta S,\nonumber\\
\Delta S &=&\intsigt \left(\kappa_{1}(D \lambda_{1}- {\bf y}\cdot \dot
{{\bf q}})+ \frac{\kappa_{2}}{2} (D \lambda_{2}- {\bf y}^{2})\right),
\eq
where $\kappa_{1,2}$ act as Lagrange multipliers. All we have done is to rename
the composite fields  ${\bf y}\cdot \dot{{\bf q}}$ and ${\bf y}^{2}$ as
$D \lambda_{1}$ and $D \lambda_{2}$. The factors of $D$ are natural since
each of these composite fields is a sum of $D$ terms.
 After this renaming, the Gaussian
integration over ${\bf y}$ can be done, and the action can be rewritten in
the following form:
\bq
S&+& \Delta S\rightarrow S_{1}+ S_{2}+ S_{3},\nonumber\\
S_{1}&=& \intsigt\left(-\frac{1}{2}\, {\bf q}'^{2}+ \frac{\kappa_{1}^{2}}
{2 \kappa_{2}}\, \dot{{\bf q}}^{2}\right),\nonumber\\
S_{2}&=& D \intsigt \left(\kappa_{1} \lambda_{1} + \frac{1}{2} \kappa_{2}
\lambda_{2}\right),\nonumber\\
S_{3}&=&\intsigt \left(i\, \bar{\psi} \dot{\psi} - D\, \tilde{g}\,
\bar{\psi}\sigma_{1}\psi +\frac{D}{2}\, \bar{\psi}\Big(\lambda_{1}
(1-\sigma_{3})-\frac{1}{2} \alpha^{2} \lambda_{2} (1+\sigma_{3})\Big)
\psi\right).\nonumber\\
& &
\eq

 In the large $D$ limit, $\kappa_{1,2}$
and $\lambda_{1,2}$ can be replaced by their vacuum expectation values:
\be
\kappa_{1}\rightarrow \kappa_{1,0}=\langle\kappa_{1}\rangle,\;\;
\kappa_{2}\rightarrow \kappa_{2,0}=\langle\kappa_{2}\rangle,\;\;
\lambda_{1}\rightarrow\lambda_{1,0}=\langle\lambda_{1}\rangle,\;\;
\lambda_{2}\rightarrow\lambda_{2,0}=\langle\lambda_{2}\rangle,
\ee
and therefore these fields become classical in this limit. In addition,
an important simplification is achieved by setting the total momentum
${\bf p}$ carried by the whole graph equal to zero:
$$
{\bf p}=\int_{0}^{p^{+}} d\sigma\,{\bf q}'=0.
$$
This configuration, which can always be reached by a suitable Lorentz
 transformation, allows us to impose the periodic boundary conditions
$$
{\bf q}(\sigma=0,\tau)={\bf q}(\sigma= p^{+},\tau).
$$
The advantage of choosing this configuration is that it is
 translationally invariant in both the $\sigma$
and the $\tau$ directions, and consequently the classical fields
$\kappa_{1,2}$ and $\lambda_{1,2}$ can be set equal to constants
independent of coordinates. It then follows that 
\be
A^{2}=\kappa_{1}^{2}/\kappa_{2}\rightarrow
\kappa_{1,0}^{2}/\kappa_{2,0}
\ee
tends to a constant in the limit of large $D$.
 We note that $S_{1}$ in eq.(19) is the standard string action,
with the slope $\alpha'$ given by
\be
\alpha'^{2}=\frac{A^{2}}{4}=\kappa_{1,0}^{2}/4 \kappa_{2,0}.
\ee
In general, this a fluctuating dynamical field, so it is far from clear
that $S_{1}$ represents a real string with a constant slope. In the large 
$D$ limit, however, since $A^{2}$ tends to a constant, so does the slope,
 and, if this constant is positive and different from zero, a real string
has formed. We will later see that this constant is never
negative; however, it could vanish. In that case, we have a zero
slope string theory, which is another name for  a field theory.
 To conclude, there is string formation only if the ground
state expectation value of $A^{2}$ is non-zero; otherwise, we have
a field theory. Therefore, the ground state expectation value of $A^{2}$
serves as an order parameter that distinguishes between the field theory
limit and string formation. In the leading large $D$ limit, this
expectation value will turn out to be non-zero if the parameters of the
model are in a suitable range, endowing the string
with a constant non-zero slope. After the corrections to the large
$D$ limit are taken into account, the string slope becomes dynamical and
it can fluctuate.

If we replace $A^{2}$ by its constant expectation value, 
 the functional integration over ${\bf q}$ in $S_{1}$ can easily
 be done, with the result
\bq
S_{1}&\rightarrow& \frac{i}{2}\,  D\, Tr\ln\left(-\partial_{\sigma}^{2} 
+ A^{2} \partial_{\tau}^{2}\right)\nonumber\\
&=& -\frac{D}{4 \pi} (\tau_{f}-\tau_{i}) \int dk \sum_{n\in Z}
\ln\left((\frac{2 \pi n}{p^{+}})^{2} + A^{2} k^{2}\right).
\eq
This needs a ultraviolet cutoff in the variable $k$ to make sense, so we
 introduce a smooth cutoff function $f(k/\Lambda)$ by letting
\be
\int dk\sum_{n}\ln\left((\frac{2\pi n}{p^{+}})^{2}+ A^{2} k^{2}\right)
\rightarrow \int dk \,f(k/\Lambda) \sum_{n}\ln\left((\frac{2\pi n}{p^{+}})^{2}
+A^{2} k^{2}\right).
\ee
This expression is not yet convergent, but the divergence is an additive
constant independent of $A^{2}$. Since the meanfield equations only
invove the derivative $S_{1}$ with respect to $A^{2}$, we can safely
make the subtraction
$$
S_{1}(A^{2})\rightarrow S_{1}(A^{2})- S_{1}(0),
$$
and arrive at the finite result
\be
S_{1}(A^{2})- S_{1}(0)= -\frac{D}{4\pi} (\tau_{f}-\tau_{i})\left(
p^{+}|A| \int dk\,k\,f(k/\Lambda)-\frac{2 \pi^{2}}{3 |A| p^{+}}\right).
\ee
In any case, if we did not drop the ghost action $S_{g}$, this additive
constant would be cancelled by the contribution from the ghost sector [5].

The integral in the first term on the right is quadratic in the cutoff;
for the sake of simplicity, we could just as well impose a sharp cutoff
and set
$$
\int dk\,k\,f(k/\Lambda)= \Lambda^{2}.
$$
In any case, a redefinition of the cutoff would yield the same result.
In the rest of the paper, we will focus only on the cutoff dependent
terms, and so, from now on, we will set
\be
S_{1}\simeq -\frac{D}{2\pi} (\tau_{f}-\tau_{i})|A|\,p^{+}\,\Lambda^{2}.
\ee
What we are doing is to study the cutoff theory prior to
renormalization. The reason for doing so is twofold: The cutoff theory
is of interest by itself; for example, in section 7, we will find
string formation for a range of the values of the coupling constant.
Also, we want to renormalize the ground state energy by introducing
a mass counter term. To do this, we have to learn about the cutoff 
dependence of various quantities by first studying the cutoff theory.

Eq.(25) could also be obtained by appealing to standard results from
string theory [8,9]. Scaling ${\bf q}$ by
$$
{\bf q}\rightarrow {\bf q}/A,
$$
the $tr\ln$ of eq.(23) is transformed into
$$
Tr\ln\left(-\frac{1}{A}\, \partial_{\sigma}^{2}+ A\, \partial_{\tau}^{2}
\right).
$$
But the calculation of this $Tr\ln$ is the same as calculating the 
ground state energy of a string with the constant background
world sheet metric given by
\be
g^{0,0}= A,\;\;g^{1,1}= 1/A,\;\;g^{0,1}=g^{1,0}=0,
\ee
and the result is the same as in eq.(25). In string theory,
the cutoff dependent term, which contributes to the energy per unit
length, is cancelled by a counter term. The finite term is the famous
Casimir term which fixes the intercept.

 Although there is this simple connection between our model
 and the standard string theory,
we would like to emphasize that there are also significant differences.
For example, the cutoff dependent term in string theory is a pure constant
and it can be dropped without disturbing the dynamics. In contrast,
 the cutoff dependent term here  is proportional to 
$A^{2}=\kappa_{1,0}^{2}/\kappa_{2,0}$, which is a dynamical quantity.
 Also, in our case,
 the coordinates $\sigma$ and $\tau$ are fixed once for all, and unlike
in string theory, there is no general reparametrization invariance.
For example, one cannot eliminate the dependence on $A$ by
 mapping the metric given by eq.(27) into
$$
g^{0,0}=g^{1,1}=1,\;\;g^{0,1}=g^{1,0}=0.
$$

We have so far introduced two different cutoffs in the $\tau$ direction;
 namely, $a'$ in eq.(11)  and $\Lambda$ in eq.(24). These are in fact
related:
 If, for example, the momentum space conjugate to $\tau$ is compactified,
the corresponding period can be identified with the cutoff $\Lambda$.
 The lattice spacing $a'$ is then related to it by
\be
a'=\frac{2\pi}{\Lambda}.
\ee

\vskip 9pt

\noindent{\bf 5. The Fermionic Action}

\vskip 9pt

In this section, we will carry out the functional integral over the
fermions in $S_{3}$, eq.(19), with $\lambda_{1}$ and $\lambda_{2}$
replaced by their coordinate independent
 expectation values, or the mean values,
 $\lambda_{1,0}$ and $\lambda_{2,0}$. To avoid
divergences, we first regulate it by discretizing
the $\sigma$ coordinate on a lattice of spacing $a$. There is then a complete
decoupling of the different lattice sites, and at each  site, we have
a two level quantum mechanics problem. Instead of working with the
action, it is easier to diagonalize the corresponding
 Hamiltonian. The total Hamiltonian
can be written as a sum of $N$ mutually commuting Hamiltonians, with
$N=p^{+}/a$:
\bq
H&=& \sum_{n} H_{n},\nonumber\\
H_{n}&=& D\left(\tilde{g} \bar{\psi}\sigma_{1}\psi- \frac{1}{2}
\bar{\psi}\Big(\lambda_{1,0}(1-\sigma_{3})- \frac{1}{2} \alpha^{2}
\lambda_{2,0}(1+\sigma_{3})\Big)\psi\right)_{\sigma=\sigma_{n}}.
\nonumber\\
& &
\eq
Acting on spin up and spin down states (dotted and solid lines),
$H_{n}$, the Hamiltonian at the site $\sigma= \sigma_{n}=n a$,
 reduces to a two by two matrix:
\be
H_{n}\rightarrow D \left(\begin{array}{cc}
\frac{1}{2} \alpha^{2} \lambda_{2,0}& \tilde{g}\\
\tilde{g}& -\lambda_{1,0}
\end{array}\right)
\ee
Diagonalizing, we have the energy levels
\be
E^{\pm}_{n}=\frac{D}{2}\left(\frac{1}{2} \alpha^{2} \lambda_{2,0}
-\lambda_{1,0} \pm \sqrt{(\frac{1}{2} \alpha^{2} \lambda_{2,0} +
\lambda_{1,0})^{2} + 4 \tilde{g}^{2}}\right).
\ee

In general, we expect the level corresponding to the minus sign
to be energetically favored;
however, we will keep both options open for the time being.

Eq.(31) gives the energy of a  fermion located at a single lattice site
$\sigma_{n}=n a$; the total fermionic energy is gotten by
multiplying this by $N=p^{+}/a$. Combining the fermionic contribution
with those coming from $S_{1}$ and $S_{2}$,
 eqs.(26) and (19), the total action is given by
\bq
S^{\pm}&=& D p^{+} (\tau_{f}-\tau_{i}) \Big(-\frac{2\pi}{a'^{2}}
|\kappa_{1,0}|/\sqrt{\kappa_{2,0}}+ \kappa_{1,0} \lambda_{1,0}+
\frac{1}{2}\kappa_{2,0} \lambda_{2,0}\nonumber\\
&-& \frac{1}{2 a} \lambda_{-} \mp \frac{1}{2 a}\sqrt{\lambda_{+}^{2}
+ 4 \tilde{g}^{2}}\Big),
\eq
where we have rewritten $A$ in terms of $\kappa$'s (eq.(21)) and defined
$$
\lambda_{\pm}=\frac{1}{2} \alpha^{2} \lambda_{2,0}\pm \lambda_{1,0}.
$$ 
Since this action is proportional to $D$, in the limit of large $D$,
 it can be evaluated using the saddle point method. This amounts to
using the equations of motion in the action. Varying $\lambda_{-}$
gives
\be
\kappa_{2,0}=\alpha^{2}(\kappa_{1,0}+\frac{1}{a}).
\ee
It is convenient to define the variable x by
$$
\kappa_{1,0}=-x/a,
$$
so that
\be
\kappa_{2,0}=\frac{\alpha^{2}}{a}(1-x).
\ee
$x$ will turn out to be positive, so we can drop the absolute
value signs from now on.
The equation of motion with respect to $\lambda_{+}$ gives
\be
\lambda_{+}^{\pm}=\pm \tilde{g}\frac{1-2 x}{\sqrt{x-x^{2}}},
\ee
and the $\pm$ signs in this equation are correlated with the $\pm$ signs
in eq.(31). Here, as well as in eq.(31), we have fixed the sign of
$\tilde{g}$ to be positive, which can always be achieved by a redefinition
of the $\pm$ signs in front of it.
 Substituting these results into eq.(32), the  corresponding energy,
related to the action (32) by
$$
S^{\pm}= -(\tau_{f}-\tau_{i}) E^{\pm},
$$
 can be written as
\be
E^{\pm}= D p^{+}\left(\frac{2\pi\,x^{\pm}}{\alpha\, a'^{2}\sqrt{a
(1-x^{\pm})}}\pm\frac{2\tilde{g}}{a} \sqrt{x^{\pm} -(x^{\pm})^{2}}
\right).
\ee
In the first term on the right, the cutoff $\Lambda$ has been
 traded for $a'$ through $\Lambda= 2\pi/a'$ (eq.(28)).
Here, $x^{\pm}$ are the values of $x$ that minimize the above
total energy  for the $\pm$ solutions.

The eigenvectors of the Hamiltonian $H_{n}$ (eq.(29)) are also of interest.
Denoting the normalized eigenvectors corresponding to $\pm$ signs of the
energy by
$$
\left(\begin{array}{c} b_{1}^{\pm}\\
b_{2}^{\pm} \end{array}\right),
$$
we have
\bq
b_{1}^{\pm}&=\mp \sqrt{\frac{1}{2}\left(1\pm \frac{\lambda_{+}}
{\sqrt{\lambda_{+}^{2}+4 \tilde{g}^{2}}}\right)}&=\mp
\sqrt{1-x^{\pm}},\nonumber\\
b_{2}^{\pm}&=\sqrt{\frac{1}{2}\left(1\mp \frac{\lambda_{+}}
{\sqrt{\lambda_{+}^{2}+ 4\tilde{g}^{2}}}\right)}&=
\sqrt{x^{\pm}}.
\eq
Let us recall the physical significance of these matrix elements:
The probabilities of having a dotted line (spin up) for  the $\pm$
solutions  are given by
\be
(b_{1}^{\pm})^{2}= 1- x^{\pm},
\ee
respectively. Similarly,
the probabilities of having a solid line (spin down) for  the $\pm$
solutions are given by
\be
(b_{2}^{\pm})^{2}=x^{\pm}.
\ee

From this probability interpretation for $x$, it follows that
\be
0\leq x^{\pm} \leq 1,
\ee
which we have already tacitly assumed. Otherwise, for example, 
eqs.(35,36) would not make sense.

From eqs.(37), it is easy to show that
\be
\frac{1}{2}\langle \bar{\psi}(1-\sigma_{3})\psi \rangle=
\langle \rho \rangle= \frac{x^{\pm}}{a}
\ee
for both $\pm$ solutions. We shall see below that (eq.(42)) $x$
is the order parameter that distinguishes between field theory
and string theory: A non vanishing $x$ signals string formation,
whereas $x=0$ corresponds to a zero slope string, which is another
name for field theory. The equation above correlates $x$ with the
expectation value of the fermionic bilinear $\bar{\psi}_{2}\psi_{2}$,
so one could as well think of this bilinear as the order parameter.
On the other hand this  bilinear is the
number operator that counts solid lines: A non vanishing expectation
value for it means that a finite proportion of the area of the
world sheet is covered by the solid lines; in other words, solid lines have
condensed on the world sheet, leading to string formation. We would
like to stress that this connection between condensation of solid
lines on the world sheet and string formation is quite robust; it is valid
independent of the approximation scheme used to compute $x^{-}$.

 In the language of Feynman graphs, condensation of solid lines
means that a single graph of asymptotically infinite order is
dominating the sum over planar graphs. It is interesting to note that
this was exactly the picture proposed in the very first papers that
attempted to deduce string formation from Feynman graphs [14,15].

The next step is to determine $x^{\pm}$ by minimizing the total
energy for each solution.  Since both terms
on the right hand side of eq.(36) are positive for the $+$ solution
and they have opposite signs for the $-$ solution, we expect that
$-$ solution represents the ground state.
 We cannot quite calculate $x^{-}$ yet, since $\tilde{g}$ will
turn out to depend on $x$, and we have first to determine this
functional dependence. This will be done in the next section, but
 since, from eq.(34)
$$
\alpha'^{2}=\frac{\kappa_{1,0}^{2}}{4 \kappa_{2,0}}=\frac{
(x^{-})^{2}}{4a \alpha^{2} (1- x^{-})},
$$
and taking $\alpha^{2}=1/a'$ (eq.(28)), then
\be
\alpha'^{2}=\frac{a' (x^{-})^{2}}{4 a (1-x^{-})}.
\ee
 Therefore, if $x^{-}$ is non-zero, we can  easily see that the slope
parameter will also be non-zero ; conversely, $x^{-}=0$ means that
the slope is zero.
In reaching this conclusion, we have assumed, as we have done throughout
this paper, that
the ratio of the two lattice spacings, $a/a'$, which is scale invariant
(eq.(17)), is finite. Since this relation between
the two cutoff parameters is essential for having
a finite slope, we would like to argue that it is required by Lorentz
invariance. In fact, if we instead allowed a more general relation, say,
$$
a'=f(a),
$$
it is easy to see that, unless $f$ is of the form
$$
f(a)= c\,a
$$
where $c$ is a constant, invariance under scaling (eq.(17)) would be
violated. This would in turn imply violation of Lorentz invariance.
As a bonus, we end up with a finite slope, with no need of renormalization.
In contrast, we shall see that the other parameter of string theory,
the intercept, is cutoff dependent and needs renormalization. We note that
 there is nothing in the problem so far that fixes the ratio $a'/a$,
and therefore, a new parameter, in addition to the coupling constant,
has to be introduced into the model. It is possible that the imposition
of full Lorentz invariance will eventually fix this parameter.

What about the $+$ solution? In this case, both terms are positive
 semidefinite, so clearly, $x^{+}=0$ minimizes the energy,
and the probability of having a solid line is zero. This is the
(trivial) starting point of standard perturbation theory; namely,
no Feynman graphs and energy equal to zero. When higher order terms
in $1/D$ are taken into account, we expect $x^{+}$ to
fluctuate and allow the formation of solid lines, thereby generating
higher order Feynman graphs. To summarize, $x$ could be non-zero only
for  the $-$ solution, leading to string formation. On the other hand,
the $+$ solution always has $x=0$, corresponding to perturbative
field theory. The ground state energy of the $+$ solution is either
greater than or equal to that of the $-$ solution.

\vskip 9pt

\noindent{\bf 6. The Interaction Vertex}

\vskip 9pt

As we have stressed earlier, following eq.(13), taking $\tilde{g}$
to be a constant amounts to neglecting the factor of $1/(2 p^{+})$
in the world sheet propagator (eq.(2)). We will now show that
this factor can be taken 
 into account in the leading mean field approximation, and as a result,
$\tilde{g}$ becomes a function of the variable $x$. We now proceed
to calculate this function. It turns out that,
for our purposes, it is more convenient to associate this factor 
$1/(2 p^{+})$ with
the vertices, rather than with the propagators. Consider two interaction
vertices, with the propagators labeled 1,2 and 3 meeting at the vertex
 as shown in Fig.4.
\begin{figure}[t]
\centerline{\epsfig{file=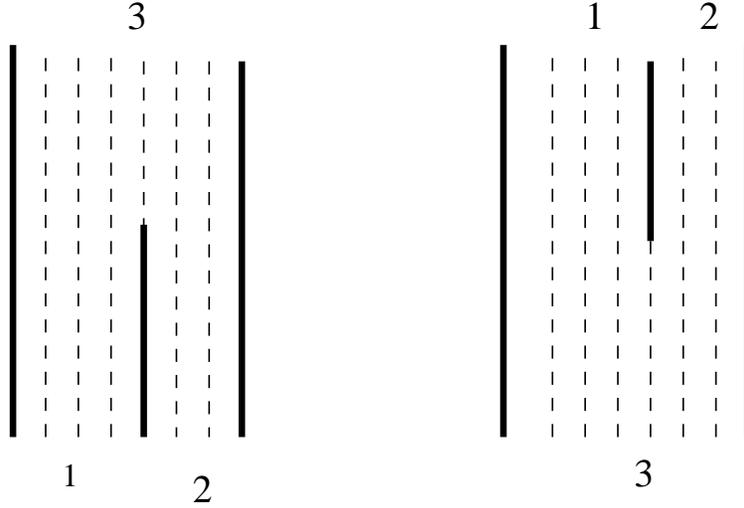,width=10cm}}
\caption{Two Vertices}
\end{figure}
 In one of them, a solid line turns into
a dotted line, and in the other, the reverse takes place. With the
first vertex, we associate a factor
\be
V_{+}= \frac{g}{8\,p_{1}^{+}\,p_{2}^{+}\,p_{3}^{+}},
\ee
and with the second vertex, a factor of
\be
V_{-}= g.
\ee
Here, $g$ is related to 
 the coupling constant of the $\phi^{3}$ interaction. It
is easy to check that this is equivalent to assigning a factor of
$1/(2 p_{i}^{+})$, $i=1,2,3$, to each of the propagators labeled 
by $i$. This assignment is not symmetrical between $V_{+}$ and $V_{-}$;
but this is not a problem since only the product
$$
V=V_{+}\,V_{-}
$$
matters. For example,
 we could interchange the roles of these two vertices, or
we could make a symmetrical assignment at the cost of introducing
square roots. For the time being, the above assignment will be convenient
to work with; later, we will show how to restore the symmetry between
$V_{+}$ and $V_{-}$.

At this point, one may wonder about the precise relationship between
$g$ and the  coupling constant of the $\phi^{3}$ interaction. Of course,
this depends on  renormalization , and
therefore, to relate the coupling constant of field theory to that
of the world sheet, one has to compare the
 renormalization schemes used in each case.
 Here we will simply treat $g$ as an effective coupling constant,
and we will not try to compare it to the field theoretic constant. It is
of interest to note that $g$ is a scale invariant constant, as contrasted
to $\tilde{g}$ (eq.(16)), so it passes at least one important test for
being a Lorentz scalar.  It is also finite (cutoff
independent), at least in the lowest order approximation. This is because,
for example, if it depended on $a$ in non-trivially, it could not be
scale invariant, since $a$ transforms under scaling (eq.(17)). So it
may be more appropriate to think of $g$ as a renormalized coupling
constant, rather than a bare one. Therefore, we have to do mass
renormalization (section 8), but we do not have to worry about coupling
constant renormalization. 

Eq.(43) refers to a vertex where each leg carries a fixed momentum
$p^{+}_{i}$. This means that  when we write down the vertex in the language
of field theory, we have to somehow express the $p^{+}$'s in terms of the
local fields. To do this exactly is a difficult problem; however, there
is a simple answer in the leading order of the meanfield approximation.
In this approximation,
we can replace the right hand side by its average value:
\be
V_{+}\rightarrow g\,
 \langle \frac{1}{8\,p_{1}^{+}\,p_{2}^{+}\,p_{3}^{+}}\rangle.
\ee
 To compute the indicated average, one has to figure
out the probability of occurence of a configuration with specified
$p^{+}$'s.
We recall from the last section that, in the leading order of the mean
field approximation, the probability of having a dotted line is given
by $1-x$ and that of having a solid line by $x$ (eqs.(38,39)).
Here $x$ is a constant independent of the coordinates, to be 
determined by minimizing the ground state energy.
\begin{figure}[t]
\centerline{\epsfig{file=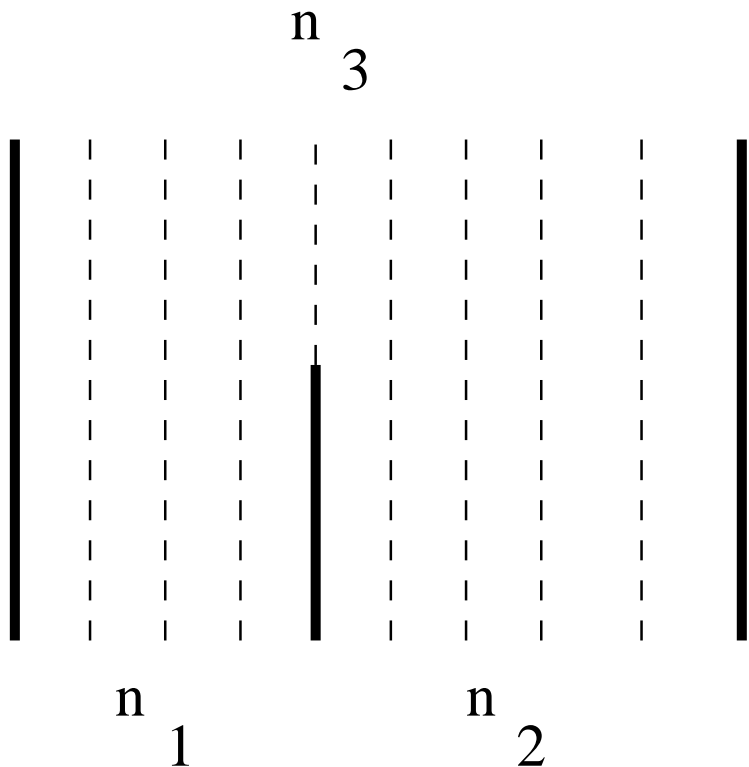,width=10cm}}
\caption{Another Vertex}
\end{figure}
 Consider the configuration
in Fig.5 of the vertex $V_{+}$, where the momenta $p_{i}^{+}$
 discretized in steps of length $a$ as usual, with  $n_{i}$ dotted lines
 associated with the propagator labeled by $i$, and
$$
p_{1}^{+}=(n_{i}+1) a,\;\;n_{3}= n_{1}+ n_{2} +1.
$$
 The probability of
having such a configuration, $P_{n_{1},n_{2}}$, depends only on the
incoming propagators 1 and 2, which completely fix the configuration.
Furthermore, in the leading order of the mean field approximation,
the probability for the occurence of a collection of solid and dotted
lines is the product of the probabilities for the occurence of each
individual line. Therefore, we have,
\be
P_{n_{1},n_{2}}= P_{n_{1}}\,P_{n_{2}},
\ee
where $P_{n}$, the probability for a single propagator is given by
\be
 P_{n}=x\,(1-x)^{n}.
\ee

The hypothesis about the factorization of probabilities used to
derive the above results 
implies lack of correlation between different lines. This is in fact the
basic hypothesis of the mean field method: To the leading order, each
line propagates independently in the background of the mean field
$x$, and the correlations between different lines  show up only
in higher order corrections.

Putting together eqs.(45,46,47), we have,
\be
V_{+}=\frac{g\,F(x)}{8},
\ee
where
\bq
F(x)&=&\sum_{n_{1}=0}^{\infty}\sum_{n_{2}=0}^{\infty}\frac{P_{n_{1},
n_{2}}}{(n_{1}+1)(n_{2}+1)(n_{3}+1)\,a^{3}}\nonumber\\
&=&\sum_{n_{1}=0}^{\infty}\sum_{n_{2}=0}^{\infty}\frac{x^{2}\,
(1-x)^{n_{1}+n_{2}}}{(n_{1}+1)(n_{2}+1)(n_{1}+n_{2}+2)\,a^{3}}.
\eq
In Appendix A, it is shown that this sum can be converted into a 
single integral. After this simplification, we have the following 
expression for $V_{+}$:
\be
V_{+}= \frac{g\,x^{2}}{(1- x)^{2}\,a^{3}}
\int_{0}^{1- x}\left( \frac{1}{y}\left(\ln(1 - y)\right)^{2}\right)d y.
\ee

At this point, it is possible to write down a fermionic interaction
term, using  $V_{+}$ and $V_{-}=g$. All we have to do is to replace
the term $\tilde{g}\,\bar{\psi}\sigma_{1}\psi$ in eq.(14) by
$$
V_{+}\, \bar{\psi}_{1}\psi_{2}+ V_{-}\,\bar{\psi}_{2}\psi_{1}.
$$
However, instead of this awkward looking non-symmetric expression,
we prefer to use a symmetrized expression. Since in calculating a
general graph, $V_{+}$ and $V_{-}$ come in pairs and always in the
form of the product $V_{-}\,V_{+}$, we are free to redefine individual
$V$'s as we wish, so long as the product remains fixed. A symmetrized
expression corresponds to the choice $V_{+}=V_{-}$. This amounts to
setting
\be
\tilde{g}^{2}=a\,V_{+}\,V_{-}.
\ee
The sudden appearence of a factor of $a$ in this equation requires
an explanation. Consider a typical solid line (Fig.3), located at
some $\sigma=\sigma_{0}$, with factors
$V_{+}$ and $V_{-}$ attached at the ends of the line. As explained
earlier, one has to integrate over the position $\sigma_{0}$ of the
coordinate. However, in deriving eq.(50) for $V_{+}$, the $\sigma$
coordinate was first latticized with a spacing $a$. Therefore, instead
of an integral, we really have a sum over the discretized positions
of the solid line. Converting this sum into an integral in the
continuum limit introduces a factor of $a$ through
$$
a\,\sum_{\sigma_{n}}\rightarrow \int d\sigma.
$$

Combining eqs.(44,50,51), we rewrite $\tilde{g}$ as
\be
\tilde{g}=\frac{g x}{a (1 - x)}\left(\int_{0}^{1 -x}
\left( \frac{1}{y} (\ln(1- y))^{2}\right) d y\right)^{1/2}.
\ee
We note that this expression for $\tilde{g}$ has the correct scaling
properties, discussed at the end of section 3. Since $g$ and $x$ are
scale invariant constants, $\tilde{g}$ scales as $1/a$ (see eq.(17)),
which is the correct result.
   
\vskip 9pt

\noindent{\bf 7. Minimizing The Ground State Energy}

\vskip 9pt

We will now rewrite the ground state energy $E^{-}$ (eq.(36)),
combining some constants to simplify the expression. We set
$$
\alpha^{2}=1/a'
$$
and define the constant $\gamma$ by
$$
\gamma= 2\pi (a/a')^{3/2}.
$$
$\gamma$  stays finite in the limit when both cutoff parameters $a$
and $a'$ go to zero, provided that the ratio $a/a'$ is kept finite.
It is then convenient to eliminate $a'$ in favor of $\gamma$ and $a$,
and to express $\tilde{g}$ in terms of $x$ through eq.(52), with the
result
\be
E^{-}=\frac{D p^{+}}{a^{2}}\left(\gamma\, \frac{x}{\sqrt{1-x}}
- 2 g h(x)\right),
\ee
where
\be
h(x)=\frac{x^{3/2}}{(1-x)^{1/2}}
\left(\int_{0}^{1 -x}\frac{1}{y} (\ln(1- y))^{2}\, d y\right)^{1/2}.
\ee

We make a couple of observations regarding this formula: Energy, an 
extensive quantity, is proportional to $p^{+}$,
 the length of the $\sigma$
interval, as it should be. It is also proportional to $1/a^{2}$, and
so it diverges in the limit $a\rightarrow 0$. This is not surprising,
since $E$ is equal to $p^{-}$, and in the frame
${\bf p}=0$ that we have chosen, the product $p^{+} p^{-}$ is equal to
the square of the mass of the (string) state. If we denote the mass of the
lowest string state by $m_{0}$, then
\be
m_{0}^{2}= p^{+}\,E^{-}.
\ee
 This is the bare mass, which
we expect to be cutoff dependent without renormalization. In the next section,
we will see how this cutoff dependence can be removed by introducing a suitable
mass counter term. Finally, note the difference in the sign of the two
terms; this is the key to the existence of a non-trivial minimum. Also,
$m_{0}^{2}$, being proportional to $(p^{+}/a)^{2}$, is scale invariant
(see eqs.(6) and (17)). Since mass is a Lorentz invariant quantity,
this is as it should be.

Before proceeding further, we note 
 one more simplification: Taking advantage of the freedom to
 renormalize the coupling constant $g$ and to redefine the cutoff
parameter $a$, we can set the constant $\gamma$ equal to unity,
so that
\be
E^{-}\rightarrow \frac{D p^{+}}{a^{2}}\left( \frac{x}
{\sqrt{1-x}}- 2 g h(x)\right).
\ee
However, $\gamma$ is not completely eliminated from the problem. For
example, the string slope $\alpha'^{2}$ still depends on
$a/a'$ and therefore on $\gamma$
(eq.(42)). The bottom line is that, whether one calls it $\gamma$ or
$a/a'$, one arbitrary constant remains in the problem.

It remains to search for the minimum of $E^{-}$ as a function of $x$.
\begin{figure}[t]
\centerline{\epsfig{file=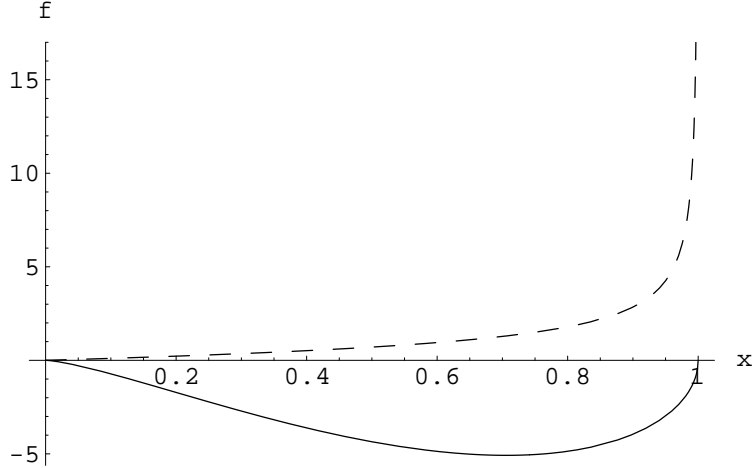,width=10cm}}
\caption{$f(x)=x/\sqrt{1-x}$ is the dashed line and $f(x)=-20\,h(x)$
is the solid line}
\end{figure}
In Fig.6, the first term in eq.(56), $x/\sqrt{1-x}$, and the function
$-20\,h(x)$, the second term for $g=10$, are seperately plotted against
$x$. Both curves start at the origin, and for small enough $g$, the
first term dominates the second term in absolute value. Therefore, for
small $g$, the minimum of $E^{-}$ is at $x=0$,
 $E^{+}=E^{-}=0$, and we have recovered the perturbative field theory
as the ground state.
Stated another way, the only solution to meanfield equations at small
coupling constant corresponds to vanishing order parameter $x$,
and therefore to the perturbative phase of the underlying field theory.
As $g$ gets bigger, there is a turning point around $g\simeq 1.3$, and past
this point, the second term dominates. The minimum $E^{-}$ now occurs
at some $x\neq 0$,  $E^{-}$ is negative at this minimum, and it
wins over $E^{+}=0$ as the ground state. Therefore, there is a critical
value of $g=g_{c}$, with $g_{c}\simeq 1.3$,
 such that for $g<g_{c}$, the system is in the
perturbative phase, and for $g>g_{c}$, it is in the string phase.
\begin{figure}[t]
\centerline{\epsfig{file=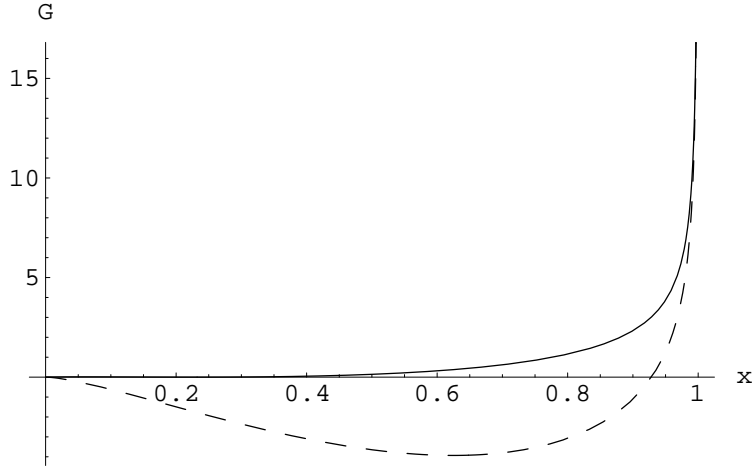,width=10cm}}
\caption{The solid line represents G(x) at g=1.3 and the dashed line
represents G(x) at g=10}
\end{figure}
 In Fig.7, the quantity
\be
G=\frac{a^{2} E_{-}}{D p^{+}}
\ee
 is plotted against $x$ for $g=1.3 $ and also for $g=10 $.
For the first value of $g\simeq g_{c}$, the
minimum is at $x=0$, and for the second one, it is at
 $x\simeq 0.625 $. As $g$ asymptotes to infinity, the location of the minimum
asymptotes to $x\simeq 7.06 $, which coincides with the location of the
minimum of $h(x)$.

To summarize, in this section, we have seen that string formation
takes place if the coupling constant is larger than a critical value.
However, it is important to realize that so far we have been talking about
an unrenormalized theory. The two physical parameters associated with a
free string are the slope and the intercept, and these should be
finite. We have already seen from eq.(42) that the slope is finite if
the ratio of the two cutoff parameters, $a/a'$ is finite as $a$ and $a'$
tend to zero. On the other hand, the intercept, which is given by
$p^{+}\,E^{-}$ (eq.(55)), diverges as $a\rightarrow 0$ because of the factor
$1/a^{2}$ in eq.(56).
 We will see in the next section that, this divergence can
be cancelled by introducing a suitable bare mass (counter)term in the
original action.

\vskip 9pt

\noindent{\bf 8.  Non-Zero Mass}

\vskip 9pt

Up to this point, we have taken the mass parameter in the propagator (eq.(2))
to be zero. It is of course important to be able to deal with non-zero
mass, since in any case, even if we set the bare mass equal to zero,
the renormalized mass will in general be different from zero. In particular,
as pointed out in the last section, the mass squared of the lowest
string state, the string intercept, is cutoff dependent. We will now show that,
by introducing a suitable mass counter term, we can eliminate this cutoff
dependence, and tune the intercept to any finite value of our choice. 
 This should be contrasted with what happens in the critical
string theory, where the intercept is fixed.

We wish to compute the contribution to the world sheet action of the
mass term in the propagator.
This calculation is greatly simplified by 
considering a thin strip of the world sheet (Fig.8),
 bounded by two lines located at
constant $\tau$ and constant $\tau+\Delta\tau$ in the $\tau$ direction,
and extending from $\sigma=0$ to $\sigma=p^{+}$ in the $\sigma$ direction.
\begin{figure}[t]
\centerline{\epsfig{file=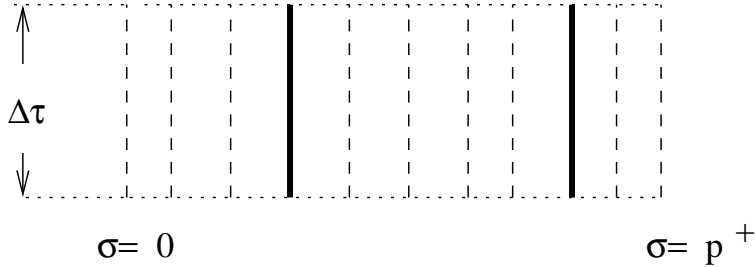,width=10cm}}
\caption{A strip of width $p^{+}$ and thickness $\Delta\tau$ on the
worldsheet}
\end{figure}
 Fig.8 shows a bunch of dotted
and solid lines in this strip, representing propagators that propagate
for an infinitesimal time interval $\Delta\tau$.  The contribution
 of the mass term to the path integral, for small $\Delta\tau$, is of the form
$$
 1+\Delta\tau\,M,
$$
 and this can be  iterated  in the $\tau$ direction to get 
$$
\exp\left((\tau_{f}-\tau_{i}) M\right),
$$
so it boils down to calculating $M$.

We will do this calculation using the mean field method, along lines
similar to the calculation of the vertex in section 6. Let the number
of dotted lines in Fig.8 be $n$, and the number of solid lines be
$N-n$, where $N$, the total number of lines, is fixed by
$$
N=p^{+}/a,
$$
where $p^{+}$ is the total width of the strip. Denote the contribution
to $M$ for a given value of $n$ by $M_{n}$, and recall that the probability
of having a dotted line is $1-x$ and that of having a solid line
is $x$. To get $M$, we weigh each configuration $M_{n}$ by the
corresponding statistical factor and add:
\be
M=\sum_{n} (1-x)^{n} x^{N- n} M_{n}.
\ee
It remains to calculate $M_{n}$, by collecting the mass dependent terms
coming from various propagators. The relevant term in eq.(2) can be rewritten
as
\be
\exp\left(-\frac{m^{2}\Delta\tau}{2(\sigma_{i+1}-\sigma_{i})}\right)
\rightarrow 1-\frac{m^{2}\Delta\tau}{2(\sigma_{i+1}-\sigma_{i})},
\ee
where $\sigma_{i}$ are the $\sigma$ coordinates of the solid lines, with
$i=1,2,...,N-n-1$. We note that, in this case, the $p^{+}$ in eq.(2)
corresponds to $\sigma_{i+1}-\sigma_{i}$, the distance between two adjacent
 solid lines. Also, we find it convenient to adopt Euclidean metric
for this calculation and therefore the factor of $i$ has been dropped.
Finally, to get $M_{n}$, one has to sum over the positions of the solid
lines
\be
M_{n}=\sum_{\sigma_{i}}\left(-\frac{m^{2}}
{2(\sigma_{i+1} -\sigma_{i})} \right).
\ee
This sum is evaluated in Appendix B, with the result
\be
M=p^{+}\frac{m^{2}\,x^{2}}{a^{2} (1-x)} \ln(x).
\ee
Comparing this to the energy with the energy in the absence of bare mass
(eq.(56)), we note the common factors
of $p^{+}$ and $1/a^{2}$, but we also see that we have to scale the mass by
$$
m^{2}\rightarrow D\,m^{2}
$$
so that the terms in the expression for the energy all have a common
factor of $D$. Otherwise, the mass term would drop out in the large 
$D$ limit. Combining eqs.(56) and (61), the total energy, including the mass,
is 
\bq
E^{-}&=&\frac{D p^{+}}{a^{2}}\Big(\frac{x}{(1-x)^{1/2}}
-2 g \frac{x^{3/2}}{(1-x)^{1/2}}\left(\int_{0}^{1-x}
\frac{\ln^{2}(1-y)}{y} dy\right)^{1/2}\nonumber\\
&-&\frac{m^{2} x^{2}}{1-x} \ln(x)\Big).
\eq
Since $x\leq 1$, the contribution of the mass term to the total
energy is positive. Remembering that the sum of the other two terms
was negative, we see that the mass term tends to raise the
ground state energy, in agreement with what one would expect.

\begin{figure}[t]
\centerline{\epsfig{file=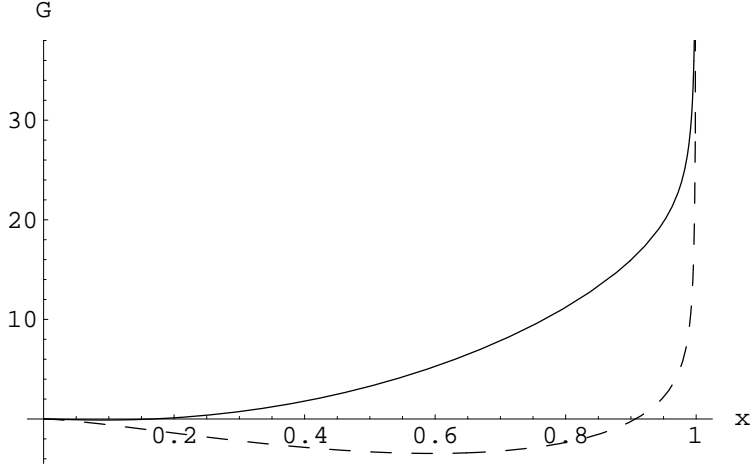,width=10cm}}
\caption{The dashed curve is G(x) for $m^{2}=1$ and the solid curve
is G(x) for $m^{2}=20$}
\end{figure}
In Fig.9, $G$ (eq.(57))
  is plotted against $x$ for $g=10 $, and for
two different values of $m^{2}$: $m^{2}=1$, the dashed curve and
$m^{2}=20$, the solid curve. 
 For $m^{2}=1 $, the curve is very similar to the one in
the massless case: There is non-trivial minimum around $x\simeq 0.6$. At the
larger value $m^{2}=20$, the curve flattens and the minimum shifts to
$x=0$. This means that string formation
takes place only if the mass is not too large, and the coupling
constant is large enough. Otherwise, the model is in the
perturbative field theory phase.

At this point, it is important to remember that so far, we have
been talking only about the cutoff dependent part of the enegy,
which is proportional to $1/a^{2}$. Similarly, the bare mass term
 makes a contribution proportional to $1/a^{2}$ to the
ground state energy. We have seen that string formation takes place
if the coupling constant is large enough and the mass is sufficiently
small so that ground state energy is negative. We hasten to add that all of 
this is before renormalization. Renormalization requires that the
cutoff dependent part of the ground state energy should be zero.
We have seen above that this can be arranged by suitably tuning
the bare mass. Now the question is, is there still string formation
even after renormalization? We have seen above that zero (cutoff
dependent) ground state energy marks the borderline between the string
and field theory phases, and so we cannot conclude anything definite
on the basis of what we have so far. To decide this question, one has to
go beyond the cutoff dependent part of the ground state energy and 
compute the finite contributions. We have already seen that the second
term on right of eq.(25) is one such contribution, but there are also
similar finite terms coming from corrections to eq.(52) for $\tilde{g}$ and 
eq.(61) for M. These calculations are rather involved and they will not
be attempted in this article, and therefore, the question of whether
there is string formation after renormalization remains open. We hope 
to return to this problem in the future.

\vskip 9pt

\noindent{\bf 9. Higher Order Contributions To $S_{1}$}

\vskip 9pt

So far, we have computed the leading term in the action in the large D
limit, which is proportional to D. The next order term is D independent,
and to compute it, one has to follow the standard prescription of the
saddle point method and expand the fields $\kappa_{1,2}$ and
$\lambda_{1,2}$ around their mean value, keeping only the quadratic
terms. The functional integrals can then in principle be done, producing
the desired term in the action. This calculation was  carried 
out in [5]; here, we will briefly  review it and also discuss its
significance and its renormalization.

We first notice that there are two different sources of higher order terms:
Those coming from $S_{1}$ and those coming from the rest of the action, such
as the fermionic sector, $\tilde{g}$ and the mass term. The contribution
coming from $S_{1}$ has a special significance: It contains the kinetic
energy term for a new degree of freedom which was not present in the original
action. The rest of the higher order contributions do not seem to have
any special significance, so we will not consider them any further.

Consider the effective action resulting from carrying out the functional
integration over ${\bf q}$ in $S_{1}$(eq.(23)):
$$
S_{1}\rightarrow\frac{i}{2} D\, Tr\ln\left(-\partial_{\sigma}^{2}+ A^{2}
\partial_{\tau}^{2}\right),
$$
where $A^{2}$ can be split into the zeroth order term $A_{0}^{2}$, which
is independent of the world sheet coordinates, plus a fluctuating term
$\Delta A^{2}$: 
\be
A^{2}=A_{0}^{2}+\Delta A^{2},\;\;
A_{0}^{2}=\frac{\kappa_{1,0}^{2}}{\kappa_{2,0}}=\frac{a'}{a}\,
\frac{x^{2}}{1-x},\;\;
\Delta A^{2}= \frac{a'}{a}\, \frac{2x -x^{2}}
{(1-x)^{2}}\,\Delta x.
\ee
  We then expand in powers of $\Delta A^{2}$
in the form of a series
\be
S_{1}= S_{1}^{(0)}+ S_{1}^{(1)} +S_{1}^{(2)}+\cdots
\ee
Since $\Delta A^{2}$ is expressible in terms $\Delta x$, this
expansion can also be converted into an expansion in powers of $\Delta x$.

The leading contribution
$$
S_{1}^{(0)}=\frac{i}{2} D\, Tr\left(-\partial_{\sigma}^{2}+ A^{2}_{0}\,
\partial_{\tau}^{2}\right)
$$
was already computed in section 4.
 Since we are expanding around a saddle point, $S_{1}^{(1)}$
vanishes. The focus of our attention here is the term second order in
$\Delta A^{2}$:
\be
S_{1}^{(2)}=-\frac{i D}{4}\,Tr\left((\partial_{\sigma}^{2}- A_{0}^{2}\,
\partial_{\tau}^{2})^{-1}\partial_{\tau}(\Delta A^{2}) \partial_{\tau}\,
(\partial_{\sigma}^{2}- A_{0}^{2} \partial_{\tau}^{2})^{-1}
\partial_{\tau}(\Delta A^{2})\partial_{\tau}\right).
\ee
This term contains both a logarithmically divergent  and also a
finite part. We will first compute the divergent part, and we will
later see that we do not need to know the finite part.

Rewriting it in momentum space, we have
\be
S_{1}^{(2)}= -\frac{i D p^{+}}{16 \pi^{2}} \int d^{2}k'\,I(k')
\Delta \tilde{A}^{2}(k') \,\Delta \tilde{A}^{2}(-k'),
\ee
where $\Delta \tilde{A}^{2}(k')$, $k'=(k'_{0}, k'_{1})$, is the Fourier
transform of $\Delta A^{2}(\tau,\sigma)$, and
\bq
I(k')&=&\int d^{2} k\,\frac{\left(4k_{0}^{2}- (k'_{0})^{2}\right)^{2}}
{\left((2k_{1}+k'_{1})^{2} - A_{0}^{2} (2 k_{0}+ k'_{0})^{2}\right)
\left((2k_{1} -k'_{1})^{2} -A_{0}^{2} (2 k_{0}- k'_{0})^{2}\right)}.
\nonumber\\
&&
\eq
In the the expression for I, we have
let $p^{+}\rightarrow\infty$ and replaced the discrete sums 
 over the variables $k_{1}$ and $k'_{1}$ by integrals. Clearly, this
is permissible when one is calculating an ultraviolet
 divergent term, which is sensitive only to the large momentum limit.

 Next, we expand I in
powers of $k'$. The zeroth order term was already included in the
calculation of $S_{1}^{(0)}$, the first order term vanishes, and 
terms with powers of $k'$ greater than two are convergent.
The logarithmic divergence comes exclusively from
the quadratic terms, given below:
\be
I\simeq \frac{i\pi}{2 A_{0}^{5}}\left(A_{0}^{2}\,(k'_{0})^{2}
- (k'_{1})^{2}\right) \int\frac{d k}{k}.
\ee

This integral is both ultraviolet and infrared divergent.
The infrared divergence is due to letting $p^{+}\rightarrow\infty$;
it can be taken care of by introducing a lower limit of roughly
$1/p^{+}$ in the  integral over k. To eliminate the ultraviolet
divergence,  the integral is cutoff at the upper limit
$k=\Lambda$, where $\Lambda$ is the same cutoff used in section 4
(eq.(25)), with the result
\be
\int\frac{d k}{k}\rightarrow \ln(\Lambda\,p^{+}).
\ee
Combining eqs.(66) and (68) and transforming back to the position space gives
\be
S_{1}^{(2)}\simeq \frac{D \ln(\Lambda\,p^{+})}{32 \pi \,A_{0}^{5}}
\intsigt \left(A_{0}^{2} \left(\partial_{\tau}( \Delta A^{2})\right)^{2}
-\left(\partial_{\sigma}( \Delta A^{2})\right)^{2}\right).
\ee

 This equation tells us that $\Delta x$ represents
a new propagating degree of freedom, with its own kinetic energy. The
promotion of a constrained field into a propagating degree of freedom
 should be familiar from other two dimensional models [16,17]. Since
$x$ is related to the fermionic bilinear $\rho$ through eq.(41),
it is reasonable to interpret this new degree of freedom as a bound
state of a pair of fermions.

We would like to say a few words about the renormalization of this result.
We can get rid of the factor of $D$ and the logarithmic factor
if we scale $\Delta A^{2}$ by letting
$$
\Delta A^{2}\rightarrow \left(\frac{16 \pi^{2}}{D\ln(\Lambda\, p^{+})}
\right)^{1/2}\,\Delta A^{2}.
$$
As a result, this term is now of order zero in the large $D$ expansion,
as opposed to terms calculated in the previous sections, which were
proportional to $D$. Also, the logarithmic divergence has dissappeared.
We note that, the finite terms which we have not calculated (see the
discussion after eq.(65)), which are
also zeroth order in $D$, will all be suppressed by this logarithmic
factor. Of course, we still expect contributions from the higher order terms
in the large $D$ expansion.

It may be of some interest to express these results by writing down
a sigma model. Since the order parameter
$x$ has now become a dynamical field, we will rename it $\chi$,
 and $x$ is now the expectation value of $\chi$.
Defining $A(\chi)$ by 
$$
A(\chi)=\left(\frac{a'}{a}\right)^{1/2}\,\frac{\chi}{(1-\chi)^{1/2}},
$$
we combine eq.(70) with eq.(62)  to form the sigma model:
\be
S_{\sigma}=\frac{D\ln(\Lambda p^{+})}{8\pi}\intsigt\left(
\frac{\dot{A}^{2}}{A} -\frac{A'^{2}}{A^{3}}-\mathcal{V}(\chi)\right).
\ee
To the leading order in $D$, the potential $\mathcal{V}$ in this
equation is given by
$$
\mathcal{V}(\chi)=-\frac{1}{p^{+}}\,E^{-}(\chi),
$$
and  $E^{-}$ is the ground state energy(eq.(62)), with the argument
$x$ replaced by $\chi$.

We note that, the slope parameter is no longer a number, but
it is now given by
\be
\alpha'^{2}=\frac{a'}{4 a}\frac{\chi^{2}}{1-\chi},
\ee
and so it becomes a fluctuating dynamical field.
 We believe that this is the crucial
difference between the fundamental strings of string theory
 and the field theory strings
of the type developed in the present work. In string theory, the slope 
is fixed, whereas here, it is a dynamical variable.
In particular, it can fluctuate and make a transition from the saddle
point $x\neq 0$ to the other saddle point $x=0$. We recall
from section 5 that this latter saddle point corresponds to perturbative
field theory.  In the present work, the world sheet configuration we
have chosen is a cylinder of infinite extent in the $\tau$ direction.
For such a configuration, and for the cutoff theory before renormalization,
we have shown that the string forming saddle point at $x\neq 0$ is 
energetically favored. However, for other configurations of the world
sheet, the other saddle point at $x=0$ may be more important. For
example, the other saddle point may contribute to a world sheet
configuration appropriate to a scattering process. What we have in mind is,
for example, a high energy and fixed angle scattering process, which
is represented by the scattering of the fundamental constituents (partons)
of a field theoretic model. It would be very nice if the saddle point
$x=0$ was dominated this process, whereas the the other saddle point,
$x\neq 0$ dominated the high energy Regge limit. This would then explain
how two different mechanisms, one underlying the ``hard'' high energy
scattering and the other underlying the ``soft'' high energy scattering,
could coexist. Inspired by the AdS/CFT correspondence, models of this type
have been costructed [18,19,20]. It is of interest to note that, in these
models also the string (Regge) slope is allowed to fluctuate.

\vskip 9pt

\noindent{\bf Conclusions}

\vskip 9pt

This article is an extension of the earlier work [2,3,4,5] on summing
planar graphs by putting them on the world sheet. Although
 as in the earlier work,
 our guinea pig theory is still the $\phi^{3}$ theory and the approximation
scheme used is still the mean field method, there is also quite a bit of
  new material. In the previous work, the prefactor that
appears in the world sheet propagator (eq.(2)) had been omitted; here, we
rectify that omisssion. Also, up to now,
 the bare mass of the field $\phi$ was taken
to be zero; in this work, we introduce a non-zero bare mass
into the problem. Prior to the introduction of the mass, a cutoff
was needed to have a well defined model, and some physical quantities,
such as the ground state energy, depended on the cutoff. With the 
introduction of a mass counter term, it becomes
for the first time possible to renormalize
the model by eliminating the cutoff dependence.

Going back to the cutoff theory, we find string formation for a
range of the values of the mass and coupling constant. For the values
of these parameters outside this range, the model goes back to the
original starting point, namely, perturbative field theory. In the
special case of vanishing bare mass, this is in agreement with the
resuts of the previous work.

In contrast to the model with cutoff, we know very little about the
renormalized model. In the future, we hope to come back to and study
it. It would be interesting to find out whether there
is string formation for any range of the parameters of the model.

Another interesting problem left open for future research is to deduce
the consequences of the promotion of the string slope into a dynamical
field. As explained at the end of the last section, this could help
connect the Regge and parton regimes of high energy scattering processes.
 
\vskip 9pt

\noindent{\bf Acknowledgements}

\vskip 9pt

This work was supported in part by the Director, Office of Science, Office 
of High Energy and Nuclear Physics, of the US Department of Energy under
Contract DE-AC03-76SF00098, and in part by the National Science 
Foundation Grant No.PHY99-07949. Part of the research leading to this
article was done while I was attending the program on QCD and Strings
at KITP, Santa Barbara. I would like to thank the organizers of this
program.

\vskip 9pt

\noindent{\bf Appendix A}

\vskip 9pt

It is useful to derive an expression for $V_{+}$ that does not
involve infinite sums. For this purpose, we define an auxilliary function
by
\be
\tilde{F}(x_{1},x_{2})=
 \sum_{n_{1}=0}^{\infty}\sum_{n_{2}=0}^{\infty}\frac{
(1-x_{1})^{n_{1}}(1-x_{2})^{n_{2}}}
{(n_{1}+1)(n_{2}+1)(n_{1}+n_{2}+2)}.
\ee
The original function can be expressed in terms of $\tilde{F}$ as
\be
F(x)=\frac{x^{2}}{a^{3}}\,\tilde{F}(x_{1}=x,
x_{2}=x),
\ee
so the problem reduces to evaluating $\tilde{F}$.

It is easy to show that $\tilde{F}$ satisfies the following differential
equation:
\bq
&&\left(2- (1-x_{1})\frac{\partial}{\partial x_{1}}
- (1-x_{2})\frac{\partial}{\partial x_{2}}\right)
\left(1 - (1-x_{1})\frac{\partial}{\partial x_{1}}\right)
\left(1- (1- x_{2})\frac{\partial}{\partial x_{2}}\right)
\tilde{F}(x_{1},x_{2})\nonumber\\
&&=\frac{1}{x_{1}\,x_{2}}.
\eq
This differential equation can be integrated partially to give
$$
\left(2- (1-x_{1})\frac{\partial}{\partial x_{1}}
- (1-x_{2})\frac{\partial}{\partial x_{2}}\right)
\tilde{F}(x_{1},x_{2})=\frac{\ln(x_{1})
\ln(x_{2})}{(1 -x_{1})(1 -x_{2})},
$$
and a further integration leads to the result
\be
\tilde{F}=\frac{1}{(1 -x_{1})(1-x_{2})}\int_{0}^
{2- x_{1} -x_{2}}\frac{d y}{y} \ln\left(1 -
\frac{y(1 -x_{1})}{2 -x_{1} -x_{2}}\right)
\ln\left(1 - \frac{y(1- x_{2})}{2 - x_{1} -x_{2}}
\right).
\ee 
Finally, substituting this result for $\tilde{F}$ in eqs.(74) and (48),
we get the following expression for $V_{+}$:
\be
V_{+}= \frac{g\,x^{2}}{(1- x)^{2}\,a^{3}}
\int_{0}^{1- x} \frac{d y}{y}\left(\ln(1 - y)\right)^{2}.
\ee

\vskip 9pt

\noindent{\bf Appendix B}

\vskip 9pt
In this Appendix, we will present a derivation of eq.(61), starting with
eqs.(58) and (60). Instead of trying to do the statistical sum of eq.(58) for a
general configuration of solid and dotted lines, we will
 first consider the simple configuration, shown in Fig.10, of one
solid line at the beginning, and all the rest dotted lines.
\begin{figure}[t]
\centerline{\epsfig{file=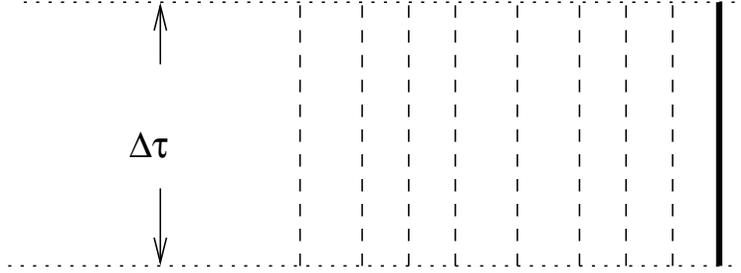,width=10cm}}
\caption{A special configuration of lines}
\end{figure}
 We will show
that the general configuration can be reached by iterating this special
configuration. Also recall from
  section 8 that we  have a thin strip of width
$\Delta\tau$ in the $\tau$ direction.
The mass contribution coming from all configurations of this type to the
 path integral is given by 
\be
\tilde{Z}(x)= \sum_{n=1}^{\infty} P_{n}(x)
\left(1- \frac{m^{2} \Delta\tau}{n a}\right),
\ee
where $n$ is the total number of lines and we have expanded to first
order in $\Delta\tau$. The mean field ansatz, eq.(47), gives
\be
P_{n}(x)=x (1-x)^{n-1},
\ee
and substituting in (78), we have
\be
\tilde{Z}(x)=1+ \frac{m^{2} x\, \Delta\tau}{a (1-x)}\ln(x).
\ee

So far, we have been summing over an arbitrary number of lines. However,
it will prove convenient to select a fixed total number $n$ of lines
from the sum. This easily accomplished by introducing a factor of
$w$ that keeps track of the number of lines, and letting
\be
K_{n}(x)\rightarrow K_{n}(w,x)=
x \,w^{n}(1-x)^{n-1}.
\ee
Eq.(80) is now replaced by
\be
\tilde{Z}(w,x)= \tilde{Z}_{0}(w,x)+ \Delta\tau\,
 \tilde{Z}_{1}(w,x),
\ee
where,
\bq
\tilde{Z}_{0}&=& \frac{w\,x}{1- w(1-x)},\nonumber\\
\tilde{Z}_{1}&=& \frac{m^{2}}{a}\,
 \frac{x\, \ln(1- w(1-x))}{1-x}.
\eq
To isolate the contribution coming from a configuration with $n$
lines, one has to expand in powers of w and pick the coefficient of
$w^{n}$.

Now consider a general configuration of lines, such as in Fig.8.
 Such a general configuration can be built from the special
configuration discussed above (Fig.10) as follows: First iterate the
special configuration as a geometric series
$$
\sum_{n=0}^{\infty}\left(\tilde{Z}(w,x)\right)^{n}=
\frac{1}{1- \tilde{Z}(w,x)},
$$
and then add to this a sum over arbitray number of dotted lines given
by
$$
\sum_{n=0}^{\infty} w^{n} (1-x)^{n}=\frac{1}{1- w (1-x)}.
$$
The result is then the contribution of the general configuration to
the path integral:
\be
Z(w,x)=\frac{1}{\left(1 -\tilde{Z}(w,x)\right)\left(
1 - w(1- x)\right)}.
\ee

We can now extract $M$ (eq.(58)) from this result as follows: First, pick
the term linear in $\Delta\tau$. And then fix the total number of lines
to be
$$
N= p^{+}/a,
$$
 by expanding in w and picking the coefficient of $w^{N}$, with
the result
\bq
M&=&\left(\sum_{n=1}^{\infty}(n-1)\left(\tilde{Z}_{0}\right)^{n-1}
\frac{\tilde{Z}_{1}}{1- w(1-x)}\right)_{w^{N}}
=\left(\frac{\tilde{Z}_{0} \,\tilde{Z}_{1}}{\left(1- \tilde{Z}_{0}\right)^{2}
\left(1- w(1-x)\right)}\right)_{w^{N}}
\nonumber\\
&=&\frac{m^{2}x^{2}}{a(1-x)}\left(\frac{w \ln(1- w(1-x))}
{(1 -w)^{2}}\right)_{w^{N}}.
\eq
Expanding in powers of w gives
\be
\left(\frac{w \ln(1- w(1-x))}{(1- w)^{2}}\right)_{w^{N}}=
\sum_{n=1}^{N-1}\left(1- \frac{N}{n}\right)(1-x)^{n}.
\ee
This result can be simplified by noticing that as $a\rightarrow 0$,
$N\rightarrow \infty$. Therefore, the first term in parenthesis on the
right hand side is negligible compared to the second term, which is
 proportional
to N. Also, the upper limit of the sum can be changed from $N-1$ to
$\infty$.  Therefore,  as $N\rightarrow \infty$,
\be
\left(\frac{w \ln(1- w(1-x))}{(1- w)^{2}}\right)_{w^{N}}
\rightarrow -N \sum_{n=1}^{\infty}\frac{1}{n} (1-x)^{n}
=N \ln(x).
\ee
Substituting this, with $N=p^{+}/a$, in eq.(85) for M gives eq.(61),
which was to be derived.

\vskip 9pt

\noindent{\bf References}
\begin{enumerate}
\item K.Bardakci and C.B.Thorn, Nucl.Phys. {\bf B 626} (2002) 287,
hep-th/0110301.
\item K.Bardakci and C.B.Thorn, Nucl.Phys. {\bf B 652} (2003) 196,
hep-th/0206205.
\item K.Bardakci and C.B.Thorn, Nucl.Phys. {\bf B 661} (2003) 235,
hep-th/0212254.
\item K.Bardakci, Nucl.Phys. {\bf B 667} (2004) 354, hep-th/0308197.
\item K.Bardakci, Nucl.Phys. {\bf B 698} (2004) 202, hep-th/0404076.
\item C.B.Thorn, Nucl.Phys. {\bf B 637} (2002) 272, hep-th/0203167.
\item S.Gudmundsson, C.B.Thorn and T.A.Tran, Nucl.Phys. {\bf B 649}
(2003) 3, hep-th/0209102.
\item C.B.Thorn and T.A.Tran, Nucl.Phys. {\bf B 677} (2004) 289,
hep-th/0307203.
\item G.'tHooft, Nucl.Phys. {\bf B 72} (1974) 461.
\item J.M.Maldacena, Adv.Theor.Math.Phys. {\bf 2} (1998) 281,
hep-th/9711200.
\item O.Aharony, S.S.Gubser, J.Maldacena, H.Ooguri and Y.Oz,
Phys.Rep. {\bf 323} (2000) 183, hep-th/9905111.
\item C.B.Thorn, Nucl.Phys. {\bf B 699} (2004) 427, hep-th/0405018.
\item For a recent review of the large N method, see M.Moshe and
J.Zinn-Justin, Phys.Rept. {\bf 385} (2003) 69, hep-th/0306133.
\item B.Sakita and M.A.Virasoro, Phys.Rev.Lett. {\bf 24} (1970) 1146.
\item D.B.Fairlie and H.B.Nielsen, Nucl.Phys. {\bf B 20} (1970) 637.
\item D.Gross and A.Neveu, Phys.Rev. {\bf D 10} (1974) 3235.
\item A.D'Adda, M.Luscher and P.Di Vecchia, Nucl.Phys. {\bf B 146}
(1978) 63.
\item J.Polchinski and M.J.Strassler, hep-th/0003136, JHEP {\bf 0305},
(2003) 012, hep-th/0209211, Phys.Rev.Lett. {\bf 88} (2002) 031601,
hep-th/0109174.
\item R.C.Brower and C.I.Tan, Nucl.Phys. {\bf B 662} (2003) 393,
hep-th/0207144.
\item O.Andreev and W.Siegel, hep-th/0410131.
\end{enumerate}

\end{document}